\documentclass[11pt]{article}
\usepackage{times}
\usepackage{geometry}
\usepackage[utf8]{inputenc}
\usepackage{enumitem,amssymb}
\newlist{thematic}{itemize}{8}
\setlist[thematic]{label=$\square$}
\usepackage{pifont}

\usepackage{fancyhdr}
\usepackage[USenglish]{babel}
\usepackage[utf8]{inputenc}
\usepackage[T1]{fontenc}
\usepackage{epsfig}
\usepackage{wrapfig}
\usepackage{color}

\usepackage[vflt]{floatflt}
\usepackage{longtable}
\usepackage{caption}
\usepackage{jheppub}   
\usepackage[dvipsnames,svgnames,x11names]{xcolor}
\usepackage[all]{hypcap} %
\usepackage{amssymb,amsmath}
\usepackage{longtable}
\usepackage{amssymb}
\usepackage{amsmath, xspace}
\usepackage{graphics,graphicx} 
\usepackage{rotating}
\usepackage{color}
\usepackage{xcolor}
\usepackage{multirow}
\usepackage{subfigure}
\usepackage{sectsty}
\usepackage[outercaption]{sidecap}
\sidecaptionvpos{figure}{c}

\usepackage{enumitem}
\setlist[enumerate]{itemsep=0pt, parsep=0pt}
\setlist[itemize]{itemsep=0pt, parsep=0pt}

\usepackage[many]{tcolorbox}
\usepackage{colortbl}
\definecolor{Myred}{rgb}{0.82,0.07,0.28}
\definecolor{Myblue}{rgb}{0.0, 0.18823529411764706, 0.28627450980392155}
\definecolor{MyLightblue}{rgb}{0.0, 0.3843137254901961, 0.6235294117647059}

\usepackage{xspace}
\def\m87{M87$^*$\xspace}
\def\sgra{Sgr~A$^*$\xspace}

\usepackage{titlesec} %
\titlespacing*{\section}{0pt}{1.0ex}{0.5ex}

\usepackage[numbers]{natbib}
\usepackage[olditem,oldenum]{paralist}
\usepackage{hyperref}
\hypersetup{
    colorlinks=true,
    allcolors=MyLightblue
}
\makeatletter
\def\NAT@anchor#1#2{%
 \hfilneg\hyper@natanchorstart{#1\@extra@b@citeb}%
 [#2]\hspace{-5pt}
 \hyper@natanchorend
}%
\makeatother
\renewenvironment{thebibliography}[1]{%
  \setlength{\parindent}{0cm}
  \par\let\par\relax
  \inparaenum}{\endinparaenum}

\begin{document}
\raggedright
\huge
\noindent Expanding the Horizon of Black Hole Imaging with AtLAST
\linebreak
\bigskip
\normalsize

\bigskip

\textbf{Authors:} 
Kazunori Akiyama (kazunori.akiyama@hw.ac.uk, Heriot-Watt University, UK);
Mariafelicia De Laurentis (mariafelicia.delaurentis@unina.it, University of Naples Federico II, Italy);
Ziri Younsi (z.younsi@ucl.ac.uk, MSSL, University College London, UK);
Yuto Akiyama (f001wbw@yamaguchi-u.ac.jp, Yamaguchi University, Japan); Dominic W. Pesce (dpesce@cfa.harvard.edu; Center for Astrophysics $|$ Harvard \& Smithsonian, USA);
Geoffrey C. Bower (gbower@asiaa.sinica.edu.tw;Academia Sinica Institute of Astronomy and Astrophysics, Taiwan); 
Kazuhiro Hada (hada@nsc.nagoya-cu.ac.jp, Nagoya City University, Japan);
Jens Kauffmann (jkauffma@mit.edu, Massachusetts Institute of Technology Haystack Observatory, USA);
Shoko Koyama (skoyama@create.niigata-u.ac.jp, Niigata University, Japan);
Kotaro Moriyama (moriyama@iaa.es, Instituto de Astrofísica de Andalucía, IAA-CSIC, Spain);
Derek-Ward Thompson (dward-thompson@lancashire.ac.uk, University of Lancashire, UK);
\linebreak

\textbf{Science Keywords:} 
stars: black holes, galaxies: active, galaxies: nuclei 
\linebreak

\captionsetup{labelformat=empty}
\begin{figure}[h]
   \centering
    \includegraphics[width=.9\textwidth]{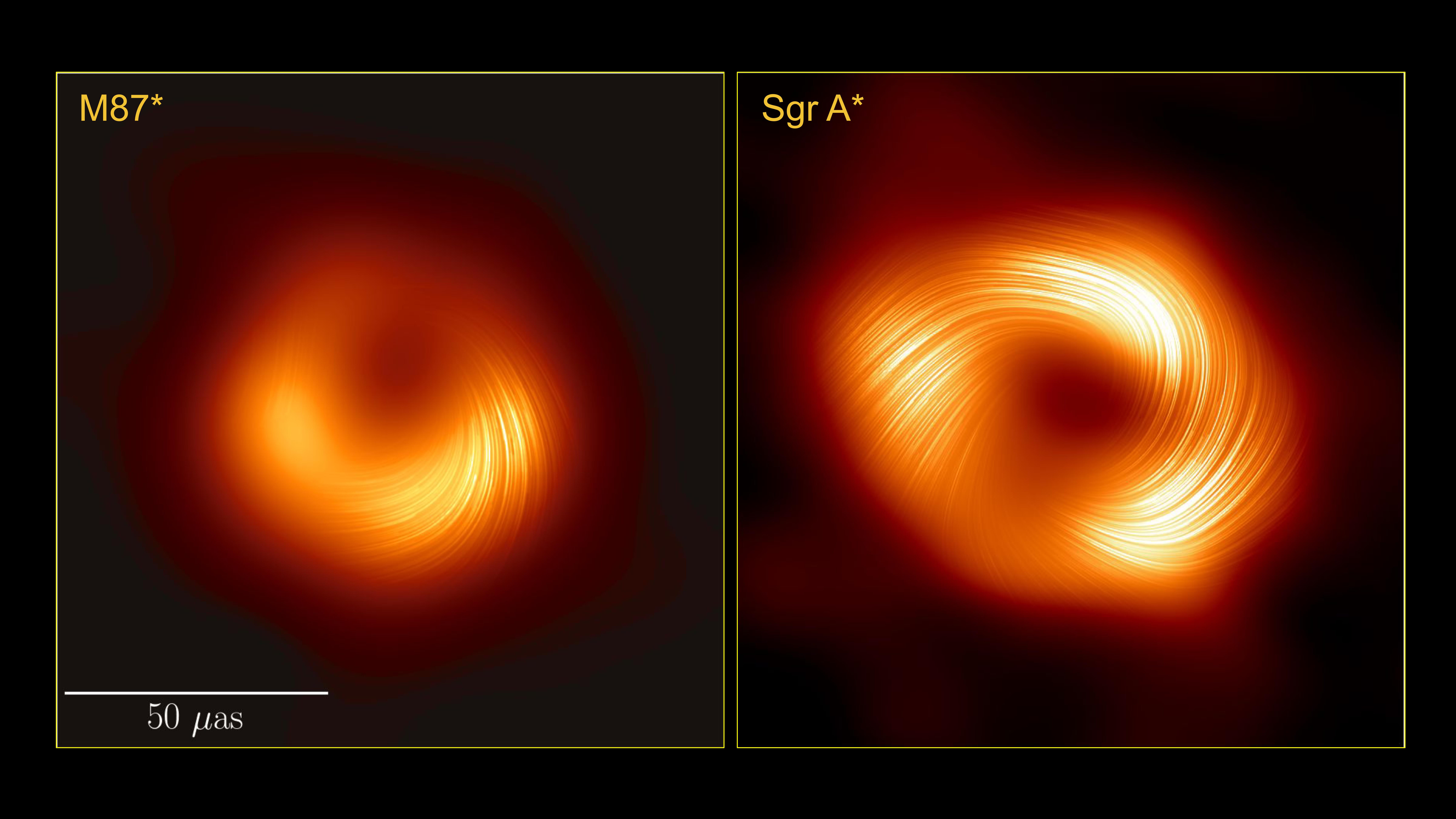}
   \caption{The first images of black holes captured by the Event Horizon Telescope (EHT), captivating billions of people across the globe and opening up a new window of black hole physics. This science white paper outlines the prospects for black hole physics enabled by an expanded sample of event-horizon-scale targets made possible by future EHT upgrades and extensions, including an extremely sensitive facility such as the Atacama Large Aperture Submillimeter Telescope (AtLAST).}
\end{figure}

\setcounter{figure}{0}
\captionsetup{labelformat=default}

\justify
\pagebreak

\justifying

\section*{Abstract}
The Event Horizon Telescope (EHT) has directly resolved and imaged two supermassive black holes and opening a new window on black hole physics.
However, the current array is limited to only these two brightest nearby targets.
This white paper outlines how future EHT upgrades, anchored by the Atacama Large Aperture Submillimeter Telescope (AtLAST), will enable a transformative expansion of the accessible population of event-horizon-scale sources.
By substantially improving sensitivity and multi-frequency capabilities, EHT+AtLAST will enable demographic studies of black hole growth, accretion physics, and jet launching across a wide range of masses, environments, and accretion states.

\section{Scientific context and motivation}
The Event Horizon Telescope \citep[EHT;][]{EHTC2017M87Paper2} has opened a new observational window by resolving and directly imaging the immediate vicinity of a black hole’s event horizon.
EHT has delivered the first-ever spatially resolved images of black hole shadows, illuminated by bright rings of radio emission from surrounding plasma, for two supermassive black holes (SMBHs):
Messier~87$^{*}$ (\m87) in the nearby radio galaxy M87 \citep{EHTC2017M87Paper1,EHTC2017M87Paper4,EHTC2017M87Paper7,EHTC2017M87Paper9,EHTC2018M87Paper1,EHTC2021M87Paper1},
and Sagittarius~A$^{*}$ (\sgra) at the center of the Milky Way \citep{EHTC2017SgrAPaper1,EHTC2017SgrAPaper3,EHTC2017SgrAPaper7}.
These resolved shadow images provide new and unique tests of general relativity and alternative theories of gravity \citep{EHTC2017SgrAPaper6}, as well as direct and accurate measurements of black hole masses \citep{EHTC2017M87Paper6,EHTC2018M87Paper2}.
Polarimetric observations of the emission rings reveal strongly magnetized accretion flows around both SMBHs \citep{EHTC2017M87Paper8,EHTC2017SgrAPaper8,EHTC2021M87Paper1}, which may enable the launching of relativistic jets through efficient extraction of energy from spinning black holes \citep{Blandford_2019}.
Together, the EHT images provide compelling evidence for the existence of black holes at the centers of galaxies and their role as the power sources of relativistic jets.

These scientific breakthroughs demonstrate the transformative science enabled by access to event-horizon-scale emission. 
At the same time, they have revealed severe limitations that prevent EHT from accessing a broader population of nearby SMBHs beyond the two sources \m87 and \sgra, which are among the brightest known targets at millimeter wavelengths, while most other nearby SMBHs are orders of magnitude fainter \citep{Johannsen_2012,Pesce_2022,Ramakrishnan_2023,Zhang_2025}.
Severe atmospheric effects significantly limit the phase coherence time and, consequently, the detection sensitivity of interferometric fringes, and they also prevent the formation of a ground-based array operating at higher frequencies to achieve improved angular resolution.
To overcome these challenges, EHT plans multi-frequency upgrades \citep{EHTC2024} that leverage the Frequency Phase Transfer (FPT) technique \citep[][]{Asaki1996,Dodson2009,Rioja2011,Rioja2023}, which has been demonstrated to deliver order-of-magnitude improvements in sensitivity \citep{2025AJ....169..229I,2025A&A...701A.132Z} and, as a byproduct, enables simultaneous, multi-scale measurements of full-polarization spectra and Faraday rotation.

In the era of anticipated multi-frequency upgrades to EHT, the planned Atacama Large Aperture Submillimeter Telescope \citep[AtLAST;][]{Klaassen2020,Ramasawmy2022} has strong potential to serve as a highly sensitive, multi-frequency anchor station for the upgraded array. 
This role will be critical for expanding the population of accessible event-horizon-scale targets \citep[see][for a detailed overview of VLBI capabilities with AtLAST]{Akiyama2023}. 
The unique strengths of AtLAST for VLBI include (but are not limited to):

\noindent {\bf Sensitive anchor in the Southern Hemisphere} ---
AtLAST is anticipated to provide the second-highest sensitivity, after the Atacama Large Millimeter/submillimeter Array (ALMA), among planned (sub)millimeter facilities in the 2030s.
Located in northern Chile, AtLAST offers very long intercontinental baselines, particularly in the north-south direction, to North American, European, and Pacific stations.
Such baselines were essential for resolving the black hole shadows in \m87 and \sgra with EHT \citep[e.g.][]{EHTC2017M87Paper4}.
When combined with co-located ALMA (equivalent to a $\sim$75\,m single dish), AtLAST would effectively provide extreme baseline sensitivity equivalent to a $\sim$90\,m single dish to Chile.

\noindent {\bf Advantages over ALMA} ---
as a single dish, AtLAST has intrinsic advantages over ALMA for VLBI, as it does not require phasing antennas distributed over hundreds of meters to kilometers to achieve high sensitivity.
First, AtLAST can expand the accessible population of faint targets, including next-generation horizon-scale SMBHs, that may be challenging to observe with phased arrays alone, which require either sufficiently bright targets for active phasing or nearby (within severeal degrees) bright calibrators for passive phasing.
Second, AtLAST can exploit its full 50\,m aperture across all observing frequencies through a dedicated multi-frequency receiving system, whereas ALMA’s sensitivity at individual frequencies is reduced when operating in sub-array modes.
According to the current ALMA 2030 Roadmap \citep{Carlson2020}, multi-band VLBI capabilities are expected to rely on such sub-array configurations.
The application and performance of FPT with sub-arrays may therefore be limited---potentially significantly---for weak sources due to non-identical optical paths \citep{2025AJ....169..229I,2025A&A...701A.132Z}.

This science white paper outlines the prospects for black hole physics enabled by an expanded sample of event-horizon-scale targets made possible by future EHT upgrades 
involving 
AtLAST.

\section{Science case}
\begin{figure}
\begin{tabular}{ccc}
\includegraphics[width=0.33\textwidth]{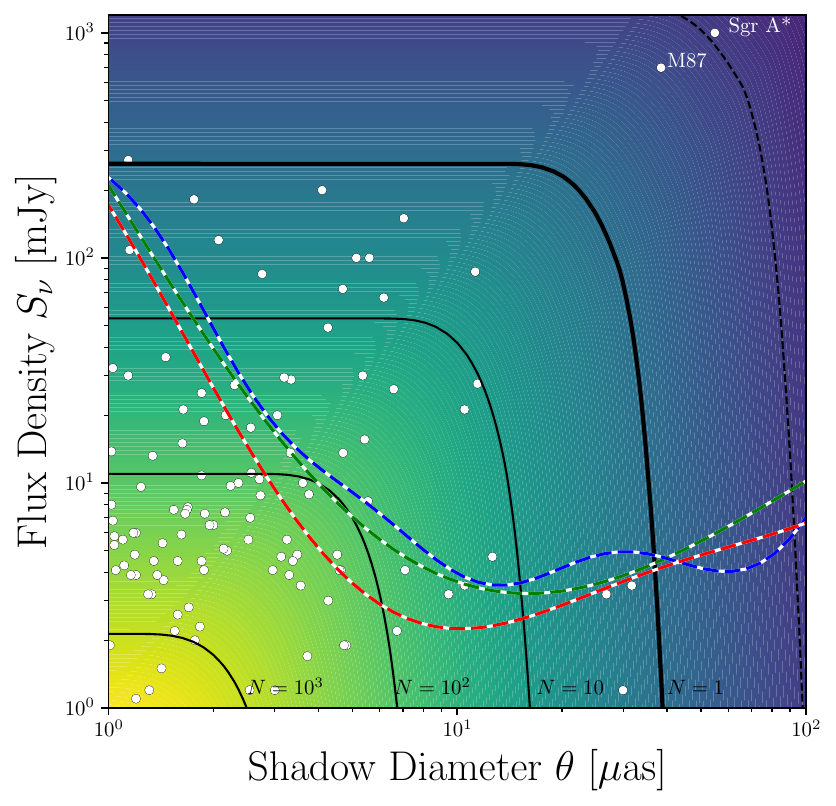} &
\includegraphics[width=0.33\textwidth]{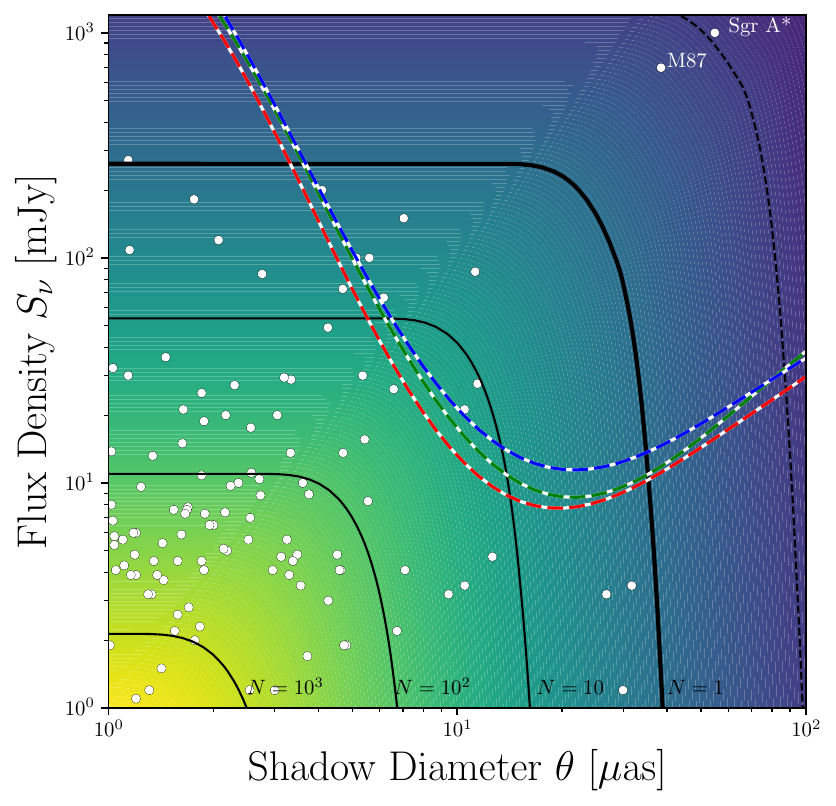} &
\includegraphics[width=0.33\textwidth]{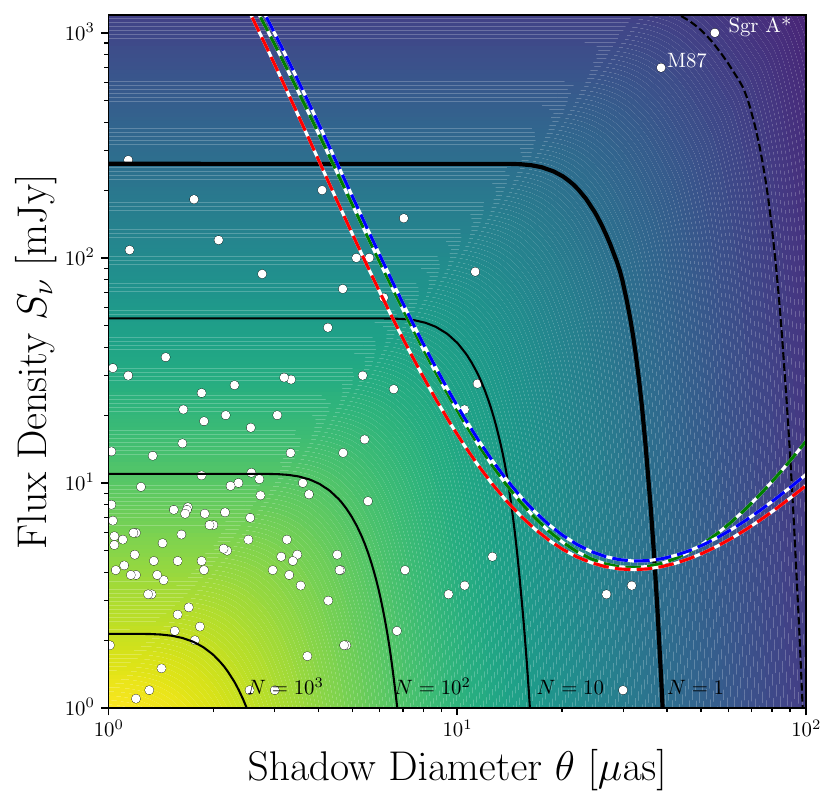}
\end{tabular}\vspace{-1em}
\caption{SMBH population studies with EHT+AtLAST in 2030s. The black contours show the estimated cumulative number density of SMBHs as a function of shadow diameter and 230 GHz flux density \citep{Pesce2021}. The white circles show the population of known SMBHs identified by the on-going ETHER survey \citep{Ramakrishnan_2023}, yet being expanded toward 2030s.
Colored lines indicate the measurement threshold values for the source size as a probe for the black hole mass (left), $\beta_2$ linear polarization polar Fourier mode as a probe for magnetic fields (middle) and the ring central depression as a probe for the black hole shadows (right). 
Those thresholds are estimated for a predicted ground-based multi-frequency EHT array in 2030s (EHT 2025+Haystack+OVRO+AMT) with synthetic EHT observations (see \cite{Pesce_2022} for methodology) for three cases: having only AtLAST (blue) or ALMA (green) in Chile, or having both two stations (red). \label{fig:thres}
}
\vspace{-1em}
\end{figure}

In \autoref{fig:thres}, we illustrate the capabilities of the EHT+AtLAST array for population studies of event-horizon-scale emission from nearby SMBHs, estimated using synthetic observations and methodology described in \cite{Pesce_2022}.
The figure shows measurement thresholds for key quantities used to infer black hole masses, magnetic field configurations, and black hole shadow properties.
Across all three metrics, AtLAST (blue curves) achieves measurement thresholds comparable to those of ALMA (green curves), indicating that AtLAST can probe a larger population of SMBHs, since these thresholds do not account for additional constraints that limit ALMA’s sensitivity to fainter targets (see Section~1).

\begin{figure}
\centering{}
\begin{tabular}{cc}
\includegraphics[height=0.3\textwidth]{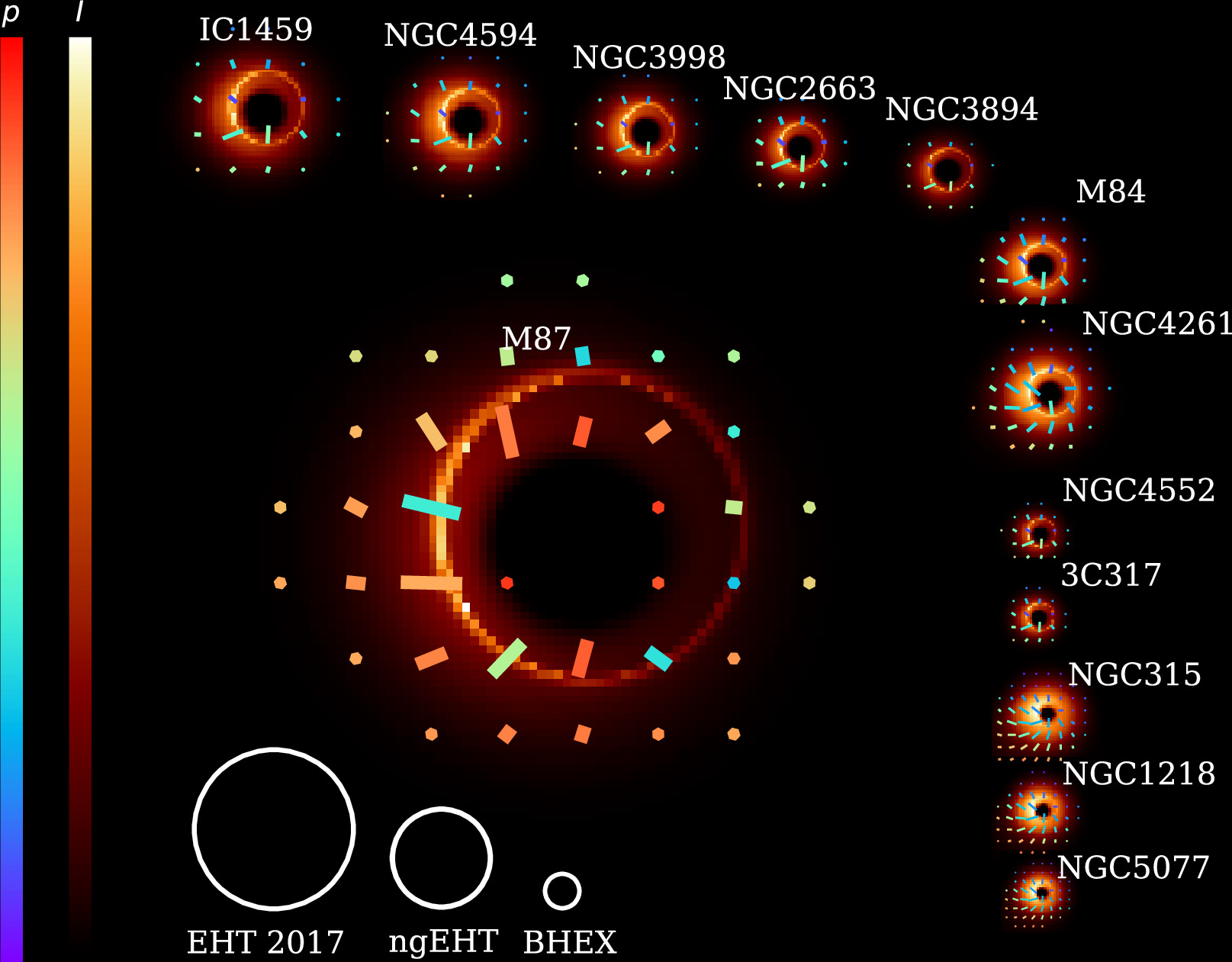} &
\includegraphics[height=0.3\textwidth]{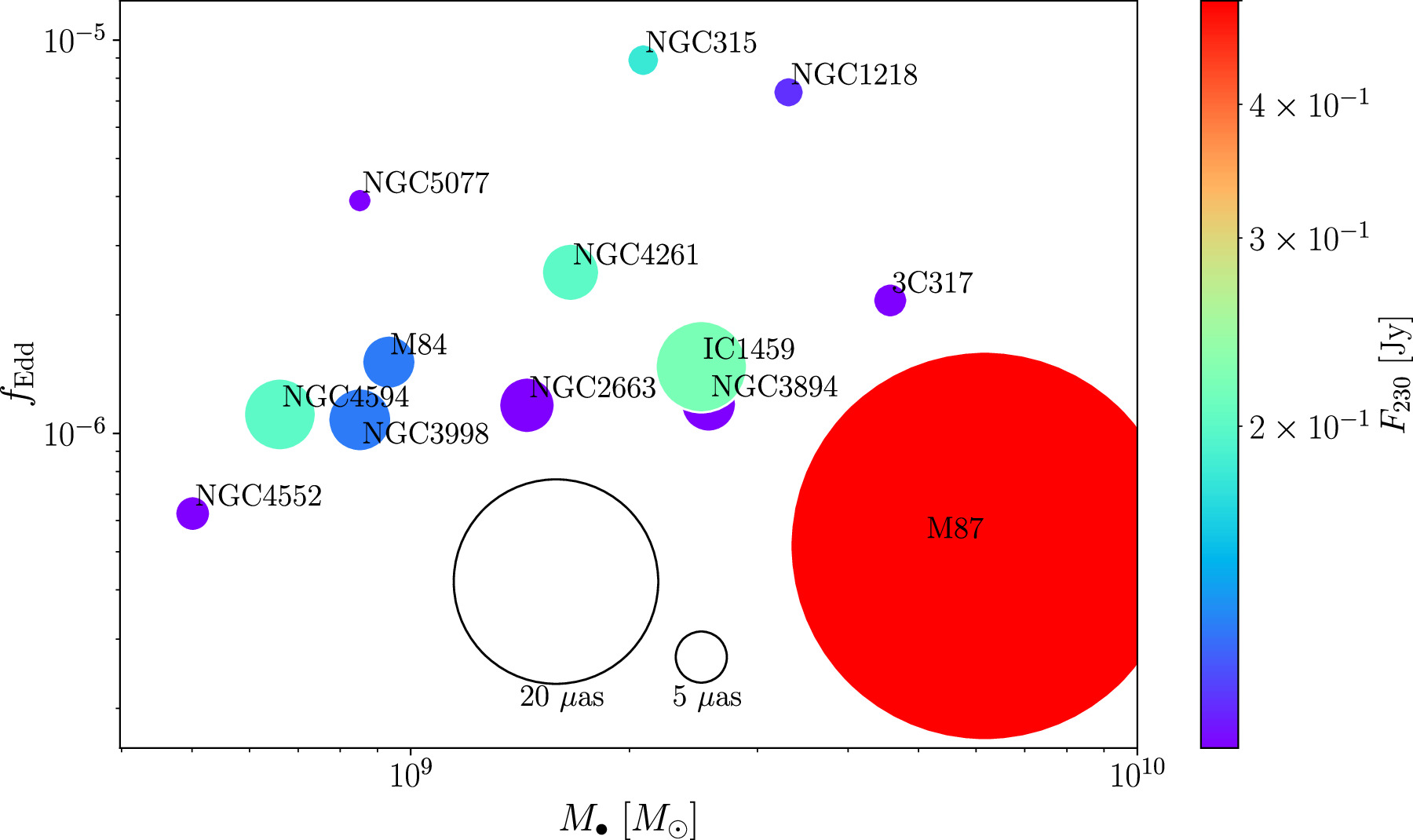} 
\end{tabular}\vspace{-1em}
\caption{Examples of nearby SMBHs for horizon-scale studies accessible with future EHT upgrades, which cover a variety of different environments 
(see \cite{Zhang_2025} for details).\label{fig:zhang25}}
\vspace{-1em}
\end{figure}

These SMBHs predominantly exhibit low Eddington ratios \citep{Ramakrishnan2023,Zhang_2025}, in contrast to most AGN samples, which are biased toward high-Eddington-ratio objects (see \autoref{fig:zhang25} for examples).
As such, they are more representative of the typical accretion states of SMBHs.
The EHT+AtLAST array will therefore enable and significantly expand demographic studies across previously unexplored regions of parameter space.
In the following, we outline a set of science questions enabled by access to this broader population of SMBHs, identified as high priorities in international community surveys \citep[e.g.][]{NAP26141}.

\noindent \textbf{How do SMBHs grow?} ---
EHT+AtLAST will map the relationships among SMBH mass, accretion rate, magnetic field structure, and potentially black hole spin, thereby linking SMBH growth to accretion states.
The sizes of horizon-scale emitting regions at (sub)millimeter wavelengths are known to provide accurate black hole mass measurements for \m87 and \sgra \citep{EHTC2017M87Paper6,EHTC2017SgrAPaper6}.
Spatially resolved linear polarization images, %
constrain accretion rates, magnetic fields, magnetizations and potentially spins of black holes \citep{EHTC2017M87Paper8,EHTC2017SgrAPaper8}.
The simulations shown in \autoref{fig:thres}, combined with estimates of the cumulative number density of SMBHs \citep{Pesce2021}, indicate that EHT+AtLAST may enable measurements of $\sim$100 source sizes and $\sim$20 linear polarization structures with detected emission rings.
The number of targets with measurable source sizes is expected to nearly double when ALMA and AtLAST jointly participate.%

\noindent \textbf{How do jets launch and accelerate?} ---
EHT+AtLAST will characterize the universality of jet-launching physics and its connection to accretion flow properties across a broad range of spacetime conditions, by imaging samples spanning a wide range of radio loudness (i.e., jet activity), viewing angles, and Eddington ratios (i.e., accretion rates).
Polarized emission is expected to vary substantially with viewing angle due to geometric effects and Faraday rotation, enabling three-dimensional insights into accretion flows and jet launching.
This capability is particularly important given that the current EHT targets, \m87 and \sgra, are both observed at relatively face-on orientations.

\noindent \textbf{How do SMBHs form, and how is their growth coupled to the evolution of their host galaxies?} ---
EHT+AtLAST will reveal SMBH properties across a diverse range of host galaxy types, including elliptical galaxies, lenticulars, and spirals (e.g., NGC~4594 in \autoref{fig:zhang25}).
Ellipticals include both central cluster galaxies like M87 and the centers of less massive galaxy groups (e.g., NGC~4261 in \autoref{fig:zhang25}).
Studies of such samples will therefore connect SMBH properties to environmental conditions across a wide range of intra- and intergalactic environments.

\section{Technical requirements and other VLBI science cases}
The key technical requirement for AtLAST VLBI is the implementation of a high-sensitivity, wide-bandwidth, multi-frequency heterodyne receiver system.
With an FPT-qualified system operating at 90, 230, and 345\,GHz, AtLAST will serve as a high-sensitivity anchor station, substantially enhancing the sensitivity and angular resolution of the EHT, its planned ground-based upgrades \citep{EHTC2024,2023Galax..11...61J}, and potential space-based extensions, including the proposed Black Hole Explorer (BHEX) mission \citep{2024SPIE13092E..2DJ,2024SPIE13092E..2EA}.
A simultaneous 230+690\,GHz receiving capability, combined with a wide-band VLBI recording system at AtLAST and a small number of other 690\,GHz-capable sites (e.g., SMA/JCMT on Mauna Kea and the GLT in Greenland), would achieve an angular resolution of $\sim$6\,$\mu$as—comparable to that of BHEX—and may enable the detection of a black hole photon ring, formed by light orbiting the black holes in \m87 and \sgra \citep{2024arXiv240701413B}.
The photon ring, currently unresolved in existing EHT images, would allow the first direct measurements of black hole spin and the most stringent tests to date of general relativity \citep{Johnson_2020,2024SPIE13092E..2DJ}.

Community white papers on (sub)millimeter VLBI \citep{2023Galax..11...61J,2013arXiv1309.3519F,2014arXiv1406.4650T,2017arXiv170504776A} further highlight a broad range of science opportunities—from faint, time-variable continuum sources in the dynamical universe to spectral-line maser sources in galactic star-forming regions and extragalactic mega-maser disks—for which AtLAST’s exceptional sensitivity will be essential, extending well beyond the science cases discussed here.

\end{document}